\documentclass[twocolumn,useAMS,usenatbib]{mn2e}
\usepackage{epsf,graphicx,amssymb}
\begin{document}
\title{Direct numerical simulations of the galactic dynamo\\ in the
  kinematic growing phase} 
\author[C. Gissinger, S. Fromang, E. Dormy]{Christophe Gissinger$^{1}$\thanks{E-mail:
    christophe.gissinger@lps.ens.fr}, Sebastien Fromang$^{2,3}$ and
  Emmanuel Dormy$^{4}$\\ 
$^{1}$LPS, CNRS UMR 8550, Ecole Normale
Sup\'erieure, 24 Rue Lhomond, 75231 Paris Cedex 05, France.\\
$^{2}$DAMTP, University of Cambridge, Centre for
Mathematical Sciences, Wilberforce Road, Cambridge, CB3 0WA, UK.\\
$^{3}$CEA, Irfu, SAp (UMR AIM), Centre de Saclay, F-91191 Gif-sur-Yvette, France.\\
$^{4}$MAG (IPGP/ENS), CNRS UMR 7154, LRA, Ecole Normale
Sup\'erieure, 24 Rue Lhomond, 75231 Paris Cedex 05, France.}
\newcommand{\Emmanuel}[1]{
[{\bf Emmanuel~:} {\sc #1}]}
\newcommand{\Christophe}[1]{
[{\bf Chritophe~:} {\sc #1}]}
\newcommand{\Sebastien}[1]{
[{\bf Sebastien~:} {\sc #1}]}

\def\bfnabla{\mbox{\boldmath $\nabla$}}

\date{\today}

\pagerange{\pageref{firstpage}--\pageref{lastpage}} \pubyear{2008}
\maketitle
\label{firstpage}

\begin{abstract}
We present kinematic simulations of a galactic dynamo model based on
the large scale differential rotation and the small scale helical
fluctuations due to supernova explosions.  We report for the first
time direct numerical simulations of the full galactic dynamo using an
unparameterized global approach.  We argue that the scale of helicity
injection is large enough to be directly resolved rather than
parameterized.  While the actual superbubble characteristics can only
be approached, we show that numerical simulations yield magnetic
structures which are close both to the observations and to the
previous parameterized mean field models. In particular, the
quadrupolar symmetry and the spiraling properties of the field are
reproduced.  Moreover, our simulations show that the presence of a
vertical inflow plays an essential role to increase the magnetic
growth rate. This observation could indicate an important role of the
downward flow (possibly linked with galactic fountains) in sustaining
galactic magnetic fields.
\end{abstract}

\begin{keywords} 
Dynamo -- Galaxies -- Supernovae -- Superbubbles.  
\end{keywords} 
 
\section{Introduction} 
It is widely accepted that magnetic fields of planets, stars and 
galaxies are generated by dynamo action, i.e., by the magnetic
field  amplification due to electromagnetic induction associated to
the  motion of an electrically conducting fluid
\citep{moffatt78}. The flow  of gas in the interstellar medium
appears to convey the essential  ingredients for such dynamo
action  \citep{wielebinski90,Rosner89}. Differential rotation in
the galactic  disk creates a strong shear along the radial
direction. This shear is  very efficient at stretching radial
magnetic field lines in the  azimuthal direction (this is known as
the $\omega$--effect).  In  combination with this large scale
effect, the turbulent motions at  small scales provide a cyclonic
flow generating poloidal magnetic  field (this is the so--called
$\alpha$--effect). Together, both  effects suggest the possibility
of an $\alpha-\omega$ type of dynamo  that might be responsible for
generating the galactic magnetic field 
\citep{parker71,Vainshtein71}. Let us note that alternative models
for  dynamo action in galaxies have been proposed through the
action of  cosmic rays \citep{hanaszetal04} or in a cosmological
context  \citep{wang&abel07}, which will not be discussed here. 
The apparent scale separation between the shear and the turbulent 
motions has often been invoked to introduce a mean field approach
for  the galactic dynamo \citep{Beck96,ferriere98}. In such
formalism, an  equation for the large scale magnetic field only is
solved, the effect  of small scales being parameterized by an
$\alpha$ term  \citep{KrauseRadler80}. Relying on mean field
equations has proven to  be a very efficient approach to the
galactic dynamo problem  \citep{ferriere92}. It is for example an
efficient way to achieve  moderate simulation time. However, the
results of mean field  simulations are intrinsically limited by
strong assumptions such as  scale separation or the statistical
properties of turbulence. It is  thus interesting to study galactic
dynamos with direct simulations of  the full problem by properly
treating the small scale flow associated  with the turbulence in
the interstellar medium and thus solve for the  magnetic field at
all scales.
\begin{figure*}
\centerline{\epsfxsize=0.9\textwidth \epsffile{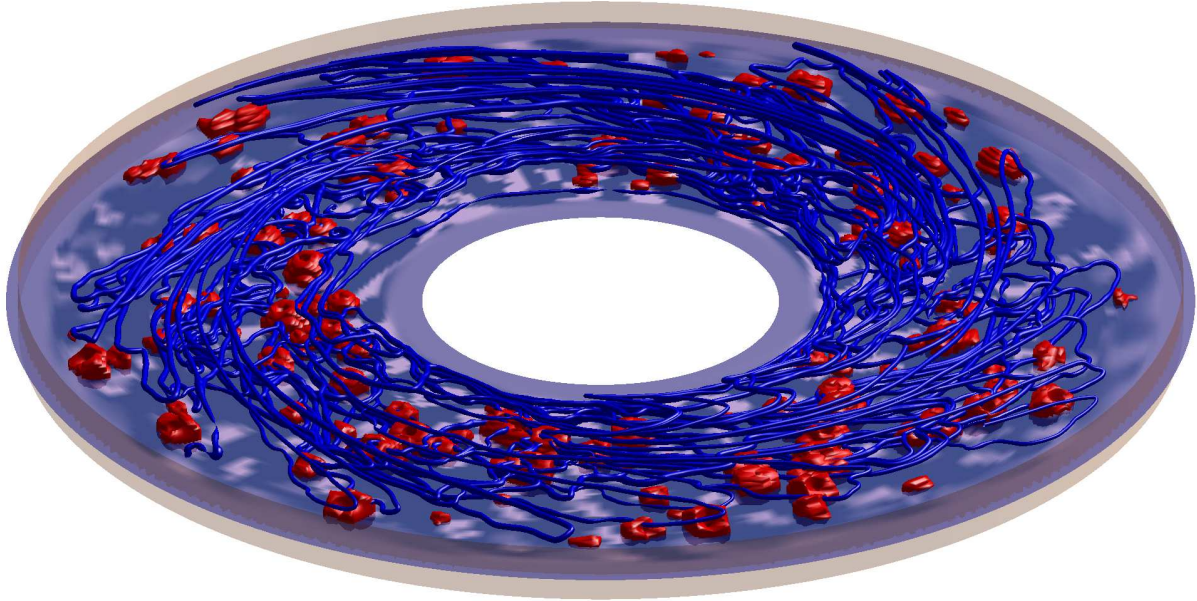} }
\vskip -2mm
\caption{Field
  lines of the magnetic field (blue) are represented for $Rm =7000$.
  Both the spiral structure of the magnetic field and its 
  quadrupolar symmetry can be identified.
  Isosurface corresponding to $1\%$ of the peak kinetic energy are
  also represented (red).}
\label{disk_azi}
\end{figure*}

It is often assumed that the most importance source of turbulence in 
the interstellar medium comes from supernova explosions 
\citep{snow&mccray79}. The positions of these explosions are not 
completely random in the disk but they often occur in cluster. This 
produces giant expanding cavities of gas known as superbubbles. These 
explosions occurring in a rotating galaxy, the expansion is affected 
by a Coriolis force. This yields cyclonic motions and thus a strong 
helicity in the gas flow \citep{ferriere98}. In such a framework, 
however, it is worth noting that the scale separation mentioned above 
is not dramatic. Superbubbles have typical sizes of the order of a few 
hundreds parsecs \citep[pc, see][]{oey&clarke97}. This is smaller, but 
not dramatically smaller than the typical vertical scale of the galaxy 
($\sim 1$~kpc). Given modern day computational resources, these 
numbers suggest that direct numerical simulations (i.e. numerical 
simulations that do not rely on an ad hoc parameterization of the 
small scales, for example through the $\alpha$--effect) are within 
reach. Indeed, \citep{gresseletal08} recently presented such 
simulations. To cope with the large resolution still needed to address 
this problem, they adopted a local approach based on the shearing box 
model. Their results indicate a good agreement of the local approach 
with mean field models.  However, the local approach they used 
precludes any global diagnostics, such as the global structure of the 
field, to be established. 
 
The purpose of this letter is to present such global numerical 
simulations, resolving the magnetic field at all relevant scales in 
the galaxy (i.e. from $100$ parsecs to $10$ kiloparsecs). To reduce 
the computational burden that would be associated with full MHD 
simulations, we work in the kinematic regime: we solve the induction 
equation using a prescribed and time dependent gas flow. The later is 
set by using an analytical velocity field which intends to reproduce 
the large scale shear associated with rotation and the effect of 
superbubble explosions on the interstellar medium.  

\section{Numerical model}
\label{numerics_sec}
The direct numerical simulations presented in this letter are fully
three dimensional. We solve the induction equation governing the
evolution of the solenoidal magnetic field $\bf B$ in a cylindrical
coordinate system $(r,\phi,z)$:
\begin{equation}
\frac{\partial {\bf B}}{\partial t} = {\bf \nabla} \times \left(
{\bf u} \times {\bf B} \right) +{Rm}^{-1} \Delta {\bf B} \, ,
\label{induction}
\end{equation}
written in dimensionless form using the advective timescale.  The
magnetic Reynolds number $Rm$ is defined as $Rm=\mu_0 \sigma r_0
U_{0}$, where $\mu_0$ is the permeability of vacuum, the typical
length scale $r_0=10 $~kpc is the radius of the galactic disk, the
typical velocity scale $U_{0}$ is the velocity of the large scale flow
(i.e. the differential rotation of the galaxy) and $\sigma$ is the
conductivity of the plasma
\footnote{We could have adopted alternative definitions of the Reynolds number, for example $Rm'=\mu_0\sigma HV_s/2$,
  where $V_s$ is the sound velocity, which is equal to the terminal
  velocity of the superbubbles (see later in the text). With this
  definition, $Rm'=Rm/200$, and the maximum value achieved in this
  work would be $Rm'=500$.}.  In our simulations the vertical extent
of the galactic disk is $H=r_0/10$, $z$ thus range from $-0.05 r_0$ to
$+0.05 r_0$. We restrict our attention here to the kinematic problem,
ignoring the back reaction of the magnetic field on the flow. The
velocity field ${\bf u}$ used in Eq.~(\ref{induction}) is analytical
and represents the differential rotation of the galaxy and the
supernova explosions. This approach also means that we do not
explicitly consider important effects such as density stratification
in the vertical direction or the induction effect which would be due
to interstellar turbulence \citep{Ruzmaikin}.

Eq.~(\ref{induction}) is solved using a finite volume approach. The 
method is described in details by \citep{teyssieretal06}: it uses
the  MUSCL--Hancok upwind method. The solenoidal character of the
magnetic  field ${\bf B}$ is maintained through the constrained
transport  algorithm \citep{yee66,evans&hawley88}.  We rely here on
the so-called  pseudo-vacuum boundary conditions for the magnetic
field. This  corresponds to imposing ${\bf B} \times {\bf n}={\bf
  0}$ at all boundaries  of the computational domain. These
boundary conditions are not fully  realistic, but they are often
used in parameterized models of  galactic dynamos and simple to
implement.  These boundary conditions are known to modify
quantitative results   (such as the threshold value for dynamo
action) but not the global qualitative solution
\citep{gissingeretal08}.  We now turn to a detailed description of the
velocity field being used. It is the sum of two terms: rotation around
the vertical axis and modification of the flow by superbubbles. In our
simulations, we use the following prescription for the rotation: ${\bf
  U}=U_0 \,{\bf e}_{\Phi}$, with a constant $U_0$. This is a good
approximation since the angular velocity is observed to be roughly
proportional to $1/r$ in galaxies. The effect of supernova
explosions is more subtle to implement. We decided to consider the
effect of superbubbles only and ignore here isolated supernovae,
as the energy input of the former is largely dominant
\citep{ferriere98}. Considering superbubbles rather than smaller
isolated supernovae yields larger scales which directly translates
into resolutions affordable with modern days computing
resources. Let us consider first the explosion of one
superbubble, in a local spherical coordinate system $(r',\theta',
\phi')$. Following the work of \citep{ferriere98}, we work under
the simplifying assumption that each explosion remnant has a
perfectly spherical shape. We thus use the simple radius evolution
law \citep{Weaveretal77}:
\begin{equation}
r'_{sb}=At^{\nu} \, ,
\label{rsb}
\end{equation}
During the expansion of each superbubble, the rotation of the galaxy 
yields a Coriolis force which tends to deflect the initially radial 
expansion and create cyclonic motions.  This is an essential step in 
classical mean field $\alpha-\omega$ description of the galactic dynamo 
\citep{ferriere98}. This Coriolis effect can be evaluated by solving 
the equation of gas motion:
\begin{equation}
\frac{\partial {\bf v}}{\partial t}={\bf F_e} - 2{\bf \Omega \times
  v_{r'}} \, ,
\label{dvdt}
\end{equation} 
where ${\bf F_e}$ is a force leading to the radial expansion described
by Eq.~(\ref{rsb}). Integrating Eq.~(\ref{dvdt}) in the radial
direction leads to the expansion~(\ref{rsb}). The azimuthal velocity
is obtained by integrating the equation (\ref{dvdt}) in the azimuthal
direction. In doing so, we made the approximation that the Coriolis
force on the superbubble is only due to the radial expansion of the
shell.  Inside the superbubble, we assume a linear variation of
velocity in radius.  An important parameter is $r'_c$, the critical
size reached by the superbubble for which the pressure in the cavity
becomes comparable to that of the surrounding medium.  At this point,
we consider that the bubble merges with the interstellar medium. This
situation generally occurs when the radial velocity of the shell
become comparable to the velocity of sound in the medium. In our
modeling, this critical velocity numerically determines the end of
existence of a superbubble. The velocity field associated with a
superbubble therefore vanishes when the radial velocity reach this
critical velocity $v_c$.  This radial expansion and the associated
Coriolis force totally determine the flow at small scales.  In most
observed galaxies, the spatial distribution of explosions in the
galaxy is rapidly decreasing away from the midplane of the disk. For
simplicity, we will assume here that all explosions occur in the
midplane only, but with random position in the disk. In actual
galaxies, there is a large observed dispersion of data about
superbubbles, yet averaged values for the explosions rate of
superbubbles are $f_0= 4.5.10^{-7}$~kpc$^{-2}$yr$^{-1}$
\citep{elmegreen85}.  Such parameters, however, are still out of reach
of present computations (especially because of the high explosion rate
which implies large numbers of superbubbles to be handled at the same
time).  We use here a lower rate of superbubbles, but more powerful
explosions, thus leading to a similar helicity input. In the
simulations reported here, $f = f_0/50$, $r'_c=0.4$, $A=0.35$ and
$\nu=0.6$. This corresponds to about $150$ superbubbles expanding in
the galactic disk at a given time in the simulations. In some cases,
we will also take into account a downward flow. Due to the simplicity
of our model, this velocity could be attributed to turbulent
diamagnetism \citep{Sokoloff} or to the galactic fountain mechanism
\citep{shapiro&field76,bregman80}. As an attempt to describe these
effects, we add the following vertical velocity to the flow:
\begin{equation}
v_{z}(z)=\frac{-\gamma
  z}{\sqrt{2\pi}2\beta} \, e^{{-z^2}/{2\beta^2}} \, .
\label{vz}
\end{equation}
It is antisymmetric with respect to the midplane and vanishes
for $z=0$. Moreover, the infall velocity decreases far away from the
midplane. The parameter $\beta$ controls the extension of the infall
region and we use here $\beta=r'_c/3$ so that the maximum of the
infall is near the region where superbubble explosions tend to
accumulate the matter.  $\gamma$ is a free parameter controlling the
amplitude of the vertical velocity. We will use here $\gamma=0.03$
throughout this paper corresponding to a typical velocity of $6$
km.s$^{-1}$.  Despite the simplifications implied by working in the
kinematic regime, large spatial resolutions are still needed in order
to correctly describe the evolution of the superbubbles at small
scales. In the runs presented here, we used a resolution of $N_r=200$,
$N_{\phi}=640$, $Nz=36$.
\section{Results}
\label{results_sec}

\subsection{General features}

\begin{figure}
\centerline{ \epsfysize=60mm \epsffile{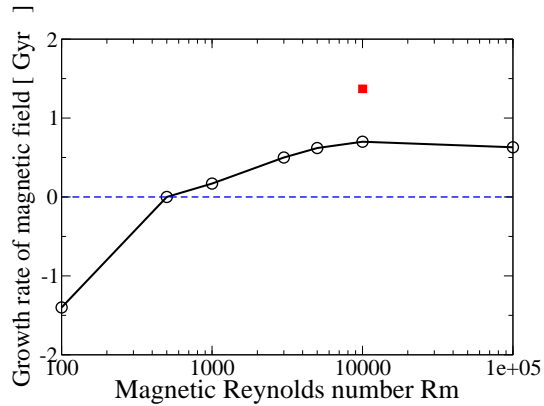} }
\caption{Growth rate of the magnetic energy as a function of the magnetic
  Reynolds number. Black circles correspond to simulations without
  infall ($\gamma=0$), whereas the filled red square corresponds to a simulation 
  with $\gamma=0.03$.}
\label{rate} 
\end{figure}
We performed seven simulations for different magnetic Reynolds number 
ranging from $Rm=100$ to $Rm=10^5$ (this would correspond to $Rm'$ 
between $0.5$ and $500$). We choose to stop the simulations after a 
few resistive times, when the growth rate of the magnetic energy is 
statistically invariant and the exponential growth is well 
established. 
 
For all of these simulations, we measure the growth 
rate of the magnetic energy. It is displayed on figure~\ref{rate} as a 
function of the magnetic Reynolds number $Rm$. It is negative when the 
magnetic Reynolds number $Rm$ is smaller than $Rm_c\sim 500$. It is positive 
for larger $Rm$, indicating exponential amplification in that 
case. For $Rm=10^5$, the growth rate is $\sigma=0.6$ Gyr$^{-1}$.  
Such growth rates are comparable to the ones obtained 
by \citep{gresseletal08}, although they seem to be larger in our 
case. 
 
The result of a typical simulation ($Rm=10^5$) once the exponentially 
growing phase is reached is illustrated in figure \ref{disk_azi} which 
shows simultaneously the structure of the magnetic field and that of 
the flow. Many superbubbles (red isosurfaces) are present at a 
given time in the model. We also show field lines (plotted in blue) 
of the magnetic field. The observed magnetic structure is the results 
of the combined effects of the superbubble explosions and the 
differential rotation of the disk. The colored slice shows the 
magnetic energy in the equatorial plane. It is strongly fluctuating 
due to the complicated nature of the flow. The overall topology of the 
magnetic field is complex. We now turn to a detailed study of its 
structure. 
 
\subsection{Structure of the magnetic field}
\begin{figure}
\centerline{ \epsfxsize=45mm \epsffile{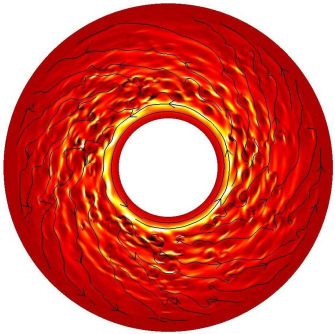} }
\vskip -2mm
\caption{Structure of instantaneous magnetic field in
  $(r,\phi)$--plane in the midplane of the galaxy at $z=0$.  Magnetic
  field lines projected in the $(r,\phi)$--plane are represented by black
  lines and the color code reflects the strength of $Bz$.}
\label{slice_turb_a} 
\end{figure}

\begin{figure}
\centerline{ \epsfxsize=45mm \epsffile{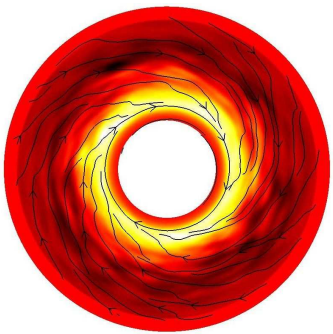} } 
\vskip -2mm
\caption{Magnetic field just before the top of the domain. Note that
the magnetic field is smooth due to the weak effect of the
superbubbles at this altitude. The sign of $B_\phi$ is reversed
compared to the midplane.}
\label{slice_turb_b} 
\vskip -4mm
\end{figure}

\begin{figure}
\centerline{ \epsfxsize=50mm \epsffile{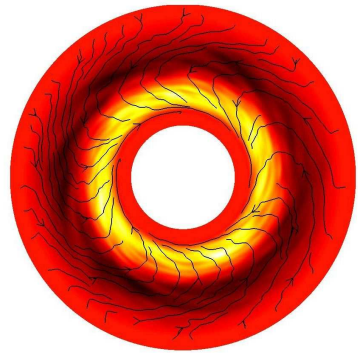} } 
\vskip -2mm
\caption{Magnetic field at $z=z0/2$. Note that $B_\phi$ change sign
when the radius is increased. }
\label{slice_turb_c} 
\end{figure}

\begin{figure}
\centerline{\raisebox{5mm}{\bf a.} \epsfxsize=86mm \epsffile{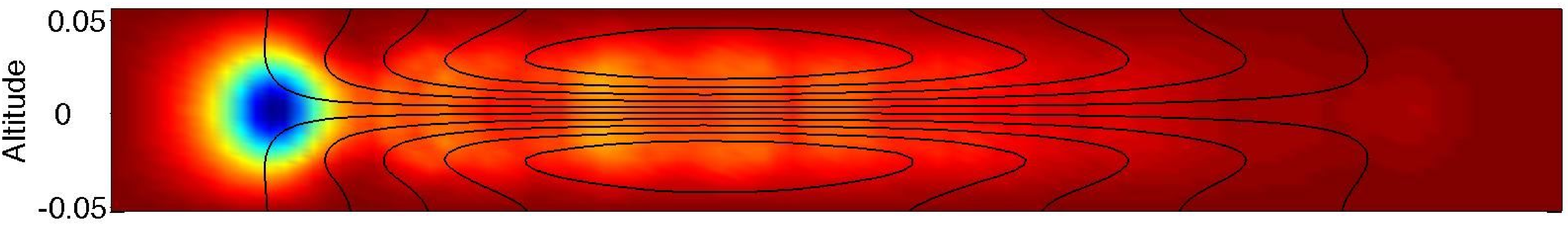} }
\vskip +1mm
\centerline{\raisebox{5mm}{\bf b.} \epsfxsize=85mm \epsffile{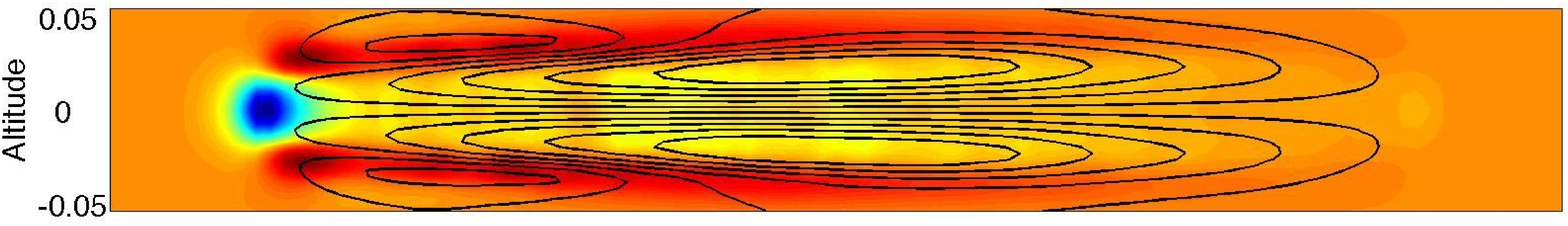} }
\vskip +2mm
\centerline{\raisebox{5mm}{\bf c.} \epsfxsize=86mm \epsffile{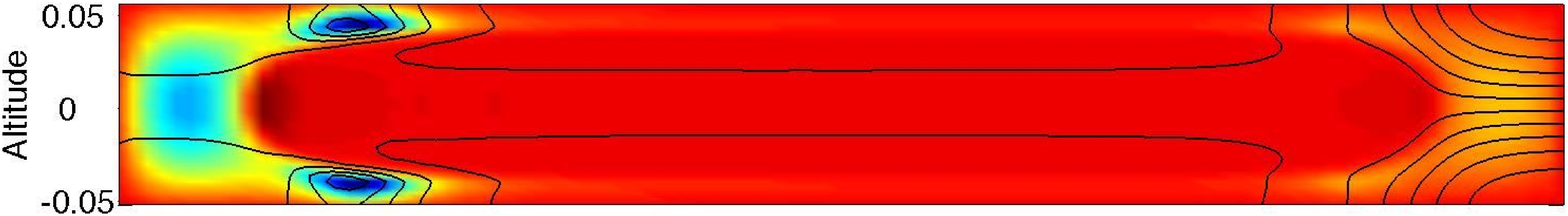} }
\centerline{\raisebox{5mm}{\bf d.} \epsfxsize=86mm \epsffile{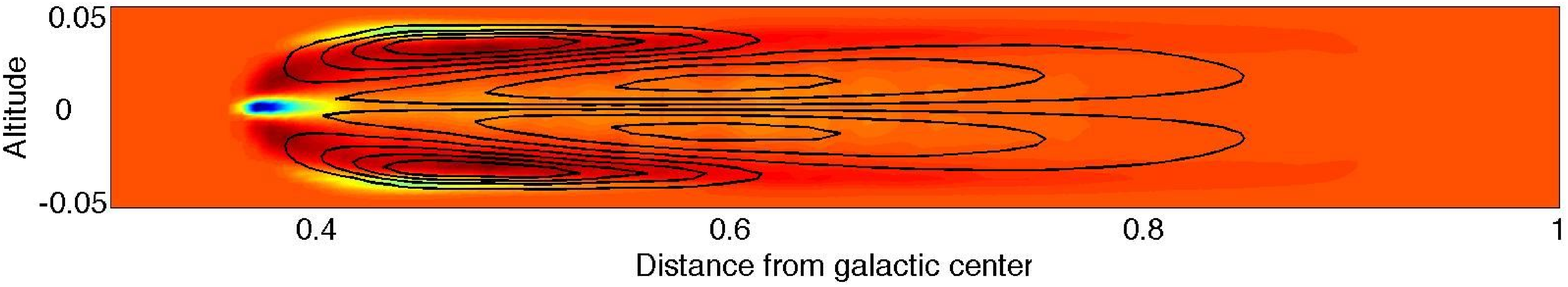} }
\caption{Magnetic structure of $\phi$-averaged magnetic field in the
  $(r,z)$--plane for: $Rm=1000$ ({\bf a.}), $Rm=3000$ ({\bf b.}) and
  $Rm=10^4$ ({\bf c.} and {\bf d.}). Flux expulsion is clearly visible
  on plot {\bf c}. Plot {\bf d.} corresponds to a simulation including
  vertical infall. Flux expulsion can thus be counteracted.}
\label{slice_rz}
\end{figure} 

The structure in the $(r,\phi)$--plane is complicated and varies with 
the altitude $z$. Figure~\ref{slice_turb_a} shows the magnetic field in 
the midplane of the galaxy (the solid lines represent field lines 
projected in this plane and the color code indicates the strength of $Bz$).  
 
In the midplane, the field is organized in a spiral structure. The 
sign of $B_{\phi}$ is constant along the radial direction. Near the 
axis of rotation of the disk, the azimuthal field largely dominates 
all others components, but rapidly goes to zero at the inner boundary 
in order to satisfy the boundary conditions. At larger radii, the 
vertical field is negligible while $B_{\phi}$ and $B_r$ are now 
comparable. Their relative value is given by the magnetic pitch angle, 
defined as $p_B={\rm atan}({B_r}/{B_{\phi}})$.  
It is remarkable that, despite the fact that numerical parameters are 
far from actual values, $p_B$ is very close to the observations: except 
near the unrealistic boundaries of the domain, the pitch angle is in 
general close to $15^o$, which is in agreement with the range 
$[-30^o,-10^o]$ observed in real galaxies \citep{shukurov07}. An 
average of the pitch angle in radius from $r=0.2$ to $r=0.8$ gives 
$p_B\simeq-15^o$. This is also in agreement with \citep{gresseletal08} 
although smaller, as they report a pitch angle around $10^o$. 
 
At higher altitudes, the structure of the field is much more 
complicated. By increasing $z$, we observe that $B_\phi$ can change 
sign. We always observe opposite sign of $B_\phi$ between the midplane 
($z=0$) and the halo ($z=+z_0$) (see figure~\ref{slice_turb_b}). At 
intermediate altitudes, $B_{\phi}$ can also reverse sign along the 
radial direction itself, as it is shown in 
figure~\ref{slice_turb_c}. 
 
While the structure in the $(r,\phi)$--plane is not very sensitive to 
the resistivity, the magnetic field in the $(r,z)$--plane presents 
different behaviors depending on the value of the Reynolds number 
$Rm$, as shown in figure \ref{slice_rz}. A quadrupolar structure is 
ubiquitous in all  
simulations but the location of the magnetic loops does depend on 
$Rm$. Indeed, the effect of superbubbles is located near the midplane 
and produces strong expulsion of magnetic field in the halo of the 
galaxy. For weak $Rm$, the magnetic resistivity counteracts this 
effect through vertical diffusion. For larger $Rm$, the weak magnetic 
resistivity cannot balance anymore the strong vertical expulsion of 
magnetic field due to multiple explosions. As a consequence, the 
quadrupole becomes unrealistically confined to the halo of the galaxy 
(Fig.~\ref{slice_rz}c), far from the active region. Although the 
creation of this external shell does not totally inhibit dynamo action, it 
 clearly decreases the magnetic growth rate. 
 
\subsection{Effect of vertical infall} 
 
This behavior indicates how the diffusion of magnetic field can play 
two opposite roles: on the one hand, it is obviously defavorable to 
dynamo action by increasing resistivity in the induction equation. On 
the other hand, diffusion can be favorable by preventing magnetic flux 
expulsion away from the midplane region where the small scale flow is 
important. However, for weak resistivity (as is the case in 
real galaxies), superbubbles expel the magnetic field out of the 
active region of galactic disk, thus inhibiting dynamo action. In that 
case, adding a vertical inflow by using Eq.~(\ref{vz}) proved to be an 
essential ingredient to dynamo action. This vertical flow indeed pumps 
the magnetic field from the halo to the midplane,  
which increases considerably the growth rate of magnetic energy. For 
$Rm=10^4$ for example, the growth rate increases from 
$\sigma=0.7$~Gy$^{-1}$ without inflow to $\sigma=1.4$~Gy$^{-1}$ with 
vertical inflow (filled red square on figure~\ref{rate}). As seen on 
figure~\ref{slice_rz}d, the magnetic field structure is again quadrupolar 
in that case and is spread out over the whole galaxy.

\section{Conclusion} 
\label{conclusion_sec} 
 
We have shown that according to our simple model, it is possible to 
perform numerical simulations of the galactic dynamo without the need 
for a mean field formalism. We thus avoid assumptions in the scale 
separation and can control more rigorously the origin of the source 
term in the induction equation. Our simulations yield magnetic field 
with two main  
characteristics: a quadrupolar symmetry in the $(r,z)$--plane and a 
roughly axisymmetric spiral configuration in the 
$(r,\phi)$--plane. Both characteristics are in good agreement with 
observations and confirm previous studies that used a mean field 
approach. A detailed study of the magnetic field topology shows a 
complicated structure, with reversals of $B_{\phi}$ along the radial or vertical 
directions. Another interesting features of the present work are the 
paradoxical role of superbubbles in the limit of very weak magnetic 
diffusion. Indeed, the turbulent flow due to explosions is, with the 
differential rotation, an essential ingredient of the $\alpha-\omega$ 
dynamo but also inhibits dynamo action by confining the magnetic field 
in the halo of the galaxy. In this context, the vertical inflow of 
interstellar gas appears as the third main 
ingredient needed for dynamo action. The downward flow observed in 
galaxies could thus be an essential mechanism of galactic dynamo 
theory. 
 
\section*{Acknowledgments} 
Computations presented in the article were performed on the IDRIS and
CEMAG computing centers. We are grateful to Romain Teyssier for
enlightening discussions and Anvar Shukurov for useful comments.
 
\bibliographystyle{mn2e} 
\bibliography{author} 
 
\end{document}